# Adjustable 3D magnetic configuration in ferrimagnetic multilayers with competing interactions visualized by soft X-ray vector tomography

J. Hermosa-Muñoz[1,2], A. Hierro-Rodríguez[1,2], A. Sorrentino[3], J. I. Martín[1,2], L. M. Alvarez-Prado[1,2], S. Rehbein[4], E. Pereiro[3], C. Quirós[1,2], M. Vélez[1,2], and S. Ferrer[3]

[1]*Depto. Física, Universidad de Oviedo, 33007 Oviedo, Spain.*
[2]*CINN (CSIC – Universidad de Oviedo), 33940 El Entrego, Spain.*
[3]*ALBA Synchrotron, 08290 Cerdanyola del Vallès, Spain.*
[4]*Helmholtz Zentrum Berlin für Materialien und Energie GmbH, 12489 Berlin, Germany*

## Abstract

Soft X-ray magnetic vector tomography has been used to visualize with unprecedented detail and solely from experimental data the 3D magnetic configuration of a ferrimagnetic $Gd_{12}Co_{88}/Nd_{17}Co_{83}/Gd_{24}Co_{76}$ multilayer with competing anisotropy, exchange and magnetostatic interactions at different depths. The trilayer displays magnetic stripe domains, arranged in a chevron pattern, which are imprinted from the central $Nd_{17}Co_{83}$ into the bottom $Gd_{12}Co_{88}$ layer with a distorted closure domain structure across the thickness. Near the top $Gd_{24}Co_{76}$ layer, local exchange springs with out-of-plane magnetization reversal, modulated ripple patterns and magnetic vortices and antivortices across the thickness are observed. The detailed analysis of the magnetic tomogram shows that the effective strength of the exchange spring at the NdCo/GdCo interface can be finely tuned by $Gd_xCo_{1-x}$ composition and anisotropy (determined by sample fabrication) and in-plane stripe orientation (adjustable), demonstrating the capability of 3D magnetic visualization techniques in magnetic engineering research.

## 1. INTRODUCTION

Advanced spintronic devices and magnetic recording media require control and characterization of multilayer magnetic systems in which the desired magnetic behaviour has to be precisely adjusted by properly tuning magnetic interactions such as exchange and magnetostatics [1, 2]. For example, ferrimagnetic materials such as Gd-Co alloys [3,4] with their adjustable magnetization offer the possibility of controlling spin-wave modes [5], spin-orbit torque, skyrmion nucleation and fast domain wall motion [6-9].

Exchange springs are very interesting features to tune the magnetic behavior of a system: when two exchange coupled magnetic layers have very different magnetic anisotropy [10], they originate characteristic domain walls at the interfaces to accommodate the reversible rotation of the soft magnetic layer under an applied field, while the hard layer stays pinned in its original magnetic configuration. They are versatile magnetic systems since they can be easily tuned adjusting the individual layer thickness and interlayer coupling strength [11-19]. Exchange springs exist in ferrimagnetic multilayers both in materials with in-plane [11, 12] and out-of-plane [13-16] magnetic anisotropy, or even crossed [17] and chiral anisotropies [18].

Most of the previous studies are based in measurements of hysteresis loops and the domain structures, which are instrumental for the understanding magnetization reversal, are inferred indirectly from comparisons with micromagnetic simulations [17]. Microscopy techniques such as Magnetic Force microscopy (MFM) can also be used to disentangle magnetization reversal at the nanoscale as has been recently reported in ferro/ferrimagnetic NiCoPt/TbFeCo but, again, with the support of micromagnetic simulations of the observed magnetic contrast [20]. In-depth characterization of magnetization profiles of interfacial domain walls can be obtained by magnetic reflectometry with either neutron [19] or X-ray based techniques [11, 21, 22] but mainly in systems with homogeneous or periodic configuration.

Full control of global and local magnetic properties in exchange coupled multilayers requires a good understanding of the 3D magnetic structures with appropriate lateral and in-depth resolutions. An initial approach based on element specific Magnetic Transmission X-ray Microscopy (MTXM) allowed to identify different magnetization reversal processes across the thickness of GdCo/NdCo/NiFe multilayers [23-25] and the physics of magnetic vortex imprinting in NiFe/CoPt nanostructures [26]. More recently, the development of Transmission X-ray Vector Tomography and Laminography [27, 28] has allowed resolving magnetization vector fields in 3D and has opened a fully experimental approach to study complex magnetization textures. These methods exploit X-ray Magnetic Circular Dichroism and differently oriented projections of the sample to be able to reconstruct its inner magnetic configuration [29, 30]. It has also been extended to dynamics allowing for 4D mapping of the magnetization [31]. MTXM has demonstrated to be excellent for revealing the nature of magnetic singularities, non-trivial magnetic topological textures and complex magnetization configurations [27, 32, 33].

In this report we have exploited the capabilities of magnetic vector tomography to determine the magnetization of a complex magnetic trilayer engineered to have competing or favoring exchange/magnetostatic interactions at different depths. The detailed analysis of the remnant magnetic configuration reveals a 3D structure that combines stripe domains, exchange spring walls across the thickness and modulated ripple patterns. In plane stripe domain orientation, adjustable through magnetic history, appears as the relevant parameter to control the strength of global interlayer coupling.

## 2. RESULTS AND DISCUSSION

### 2.1 Multilayer design

We have prepared an 80 nm $Gd_{12}Co_{88}$/80 nm $Nd_{17}Co_{83}$/80 nm $Gd_{24}Co_{76}$ multilayer designed to support an exchange spring at the top layer. Ferrimagnetic $Gd_xCo_{100-x}$ alloys present a soft magnetic behaviour with in-plane uniaxial anisotropy, whereas the stoichiometry of the central $Nd_{17}Co_{83}$ layer corresponds to a ferromagnetic alloy with weak perpendicular anisotropy (see Fig. 1(a)). The purpose of the central layer, which had the largest anisotropy in the multilayer, is to create a pattern of stripe domains with alternating up-down magnetization orientation [23] that can be used to control the configuration in the neighbouring $Gd_xCo_{100-x}$ layers via interfacial exchange and magnetostatic interactions. Also, the stripe pattern provides an interesting possibility for external adjustment of the system after sample fabrication: stripe orientation can be rotated at will with large enough in-plane magnetic fields and becomes locked at low fields by the so-called rotatable anisotropy [34, 35] which is an in-plane pseudo-uniaxial anisotropy of magnetostatic origin with easy axis along the direction of the stripes.

The magnetization of the ferrimagnetic $Gd_xCo_{100-x}$ alloy is given by the sum of the contributions of Co and Gd atom sublattices [11]. The magnetization vectors of each sublattice can be written as $\mathbf{M}(Co) = M(Co)\mathbf{m}(Co)$ and $\mathbf{M}(Gd) = M(Gd)\mathbf{m}(Gd)$, where $M(Co)$ and $M(Gd)$ are the magnetization magnitudes and $\mathbf{m}(Co)$ and $\mathbf{m}(Gd)$ are unit vectors. In these $Gd_xCo_{100-x}$ alloys, Gd and Co moments are collinear and antiparallel, so that $\mathbf{m}(Gd) = -\mathbf{m}(Co)$. The net magnetization of the alloy $\mathbf{M}(GdCo)$ is given by

$$\mathbf{M}(GdCo) = \mathbf{M}(Co) + \mathbf{M}(Gd) = (M(Co) - M(Gd))\mathbf{m}(Co) = \varepsilon\mathbf{m}(Co) \qquad (1)$$

where $\varepsilon = M(Co) - M(Gd)$ is a scalar factor that relates $\mathbf{M}(GdCo)$ with $\mathbf{m}(Co)$. The saturation magnetization of the alloy is $M_S(GdCo) = |M(Co) - M(Gd)| = |\varepsilon|$. Depending on temperature and composition, $\varepsilon$ may be either positive ($M(Co) > M(Gd)$) or negative ($M(Co) < M(Gd)$). In-plane magnetized $Gd_xCo_{100-x}$/$Gd_yCo_{100-y}$ exchange springs are usually designed by finely tuning the alloy composition so that $\varepsilon$ has opposite signs at each layer [11].

Here, the composition of the bottom $Gd_{12}Co_{88}$ layer has been chosen so that its saturation magnetization $M_S(GdCo)^{BOT}$ is large and $\varepsilon^{BOT}$ is positive at room temperature. On the contrary, the top layer $Gd_{24}Co_{76}$ composition is selected so that its compensation temperature is slightly above room temperature which implies that $M_S(GdCo)^{TOP}$ is small and $\varepsilon^{TOP}$ is negative. The sign of $\varepsilon$ was determined by Magnetooptical Transverse Kerr effect (MOTKE) hysteresis loops (see Figs. 1(b-c)), measured at room temperature with white light, on 50 nm thick $Gd_xCo_{100-x}$ control samples with the same composition as top/bottom layers in the trilayer. For incident light in the visible, MOTKE is only sensitive to Co magnetic moments [11]. Then, at positive applied fields, $\varepsilon^{BOT} > 0$ implies a positive MOTKE signal and $\mathbf{m}(Co)^{BOT}$ parallel to $\mathbf{M}(GdCo)^{BOT}$ (see Fig. 1(c)). On the contrary, $\varepsilon^{TOP} < 0$ implies negative MOTKE signal at positive applied fields since $\mathbf{m}(Co)^{TOP}$ is antiparallel to $\mathbf{M}(GdCo)^{TOP}$ (see Fig. 1(b)).

The magnetization configuration of GdCo layers will depend on three competing interactions across the sample thickness related with the magnetic parameters selected for each layer in these $Gd_{12}Co_{88}$/$Nd_{17}Co_{83}$/$Gd_{24}Co_{76}$ multilayers (see sketch in Fig. 1(a)). The first one is the exchange interaction at the interfaces, which in Rare Earth-Transition Metal alloys is dominated by $J(Co - Co)$, the exchange between Co moments [11] and favours the parallel alignment of $\mathbf{m}(Co)$ at the different layers. The second is the uniaxial anisotropy with out-of-plane easy axis at the central layer and in-plane easy axis at the outer layers. Finally, magnetostatic interactions will try to minimize magnetic charges (i.e. discontinuities in the magnetization), creating closure domains and favouring parallel alignment of $\mathbf{M}(GdCo)$ and $\mathbf{M}(NdCo)$ at the interfaces, which may either compete or cooperate with interfacial exchange depending on the sign of $\varepsilon$.

In summary, our multilayer design includes a laterally modulated domain structure (stripe domains) that can be adjusted with magnetic history, and competing exchange, anisotropy and magnetostatic interactions across the multilayer thickness. As described in what follows, vector tomography has allowed visualizing the complex domain structure demonstrating its capabilities in magnetic engineering research.

## 2.2 Soft X-ray MTXM and Vector Tomography

Figure 2 shows several MTXM images, measured at the Gd M absorption edge with different angles of incidence so that they are sensitive to the different components of the magnetic moment $\mathbf{m}(Gd)$ (angle geometry as sketched in Fig. 2). Figs. 2(a-c) correspond to a first series of images measured with $\phi = 0°$ (Tilt series 1) and Figs 2(d-f) to a second series of images measured with $\phi \approx 95°$ (Tilt series 2). A clear chevron pattern of alternating out-of-plane stripe domains is observed at normal incidence with period $\Lambda \approx 215$ nm (see Figs. 2(b) and 2(d) at $\vartheta = 0°$). In Tilt series 1 at oblique incidence (sensitive to $m_y$ and $m_z$), we observe groups of parallel stripes with similar average contrast that is reversed for opposite values of $\vartheta$ (see regions labelled as MD1-MD4 in Figs. 2(a) and 2(c) at $\vartheta = \pm 30°$). This indicates that the sample magnetization is broken into a structure of larger magnetic domains (several $\mu m$ wide) with similar average in-plane magnetization superimposed on the fast out-of-plane magnetization oscillation of the magnetic stripe pattern. These in-plane magnetization domains will be denoted as MD in the following.

However, in Tilt series 2 at oblique incidence (sensitive approximately to $m_x$ and to $m_z$), average in-plane magnetic contrast is much weaker than in Tilt series 1, with only slight differences between the pairs MD1/MD4 and MD3/MD2, indicating that the average in-plane magnetization of these MDs is mostly oriented along the $y$-axis, which is reasonable since the sample was mounted with the easy anisotropy axis of the $Gd_xCo_{100-x}$ layers approximately perpendicular to the rotation axis for Tilt series 1 (i.e. along the $y$ direction).

The 3D configuration of the magnetic moment, $\mathbf{m}(Gd)$, in the multilayer has been obtained from the reconstruction of the MTXM data sets with a vectorial tomography algorithm described in [29, 32]. Since the Co magnetic moment $\mathbf{m}(Co)$ is collinear and antiparallel with $\mathbf{m}(Gd)$ in $Gd_xCo_{100-x}$ alloys, its configuration can also be estimated from $\mathbf{m}(Gd)$ vector maps. The physics of the multilayer will be analysed in terms of the magnetic moments of the individual Gd and Co sublattices, i. e. $\mathbf{m}(Gd)$ and $\mathbf{m}(Co)$ and, depending on magnetic layer, the net magnetization $\mathbf{M}$ will either be parallel or antiparallel to each of them. We will focus on the differences in configuration between the upper and lower regions of the multilayer (i.e. GdCo layers), with a more qualitative discussion of the evolution across the thickness.

### 3.1 Out of plane magnetic moment: the role of effective magnetostatic coupling

#### 3.1.1 Analysis of experimental vector tomograms

The vector map of magnetic moments at MD1 obtained from the tomographic reconstruction reveals a complex 3D configuration as shown in Fig. 3. There is a pattern of parallel stripe domains with alternating positive/negative $m_z$ with a period $\Lambda = 215$ nm both at the top $Gd_{24}Co_{76}$ layer (Fig. 3(a)) and at the bottom $Gd_{12}Co_{88}$ one (Fig. 3(b)). However, the $m_z$ oscillation at the bottom layer is in antiphase with the oscillation at the top (see e. g. $m_z$ sign along points $P_1$ to $P_3$ marked in Figs. 3(a-b) for comparison), and its amplitude is almost three times larger at the bottom than at the top layer.

Figure 3(c) shows a cross section of the magnetic moment vector map taken along a line transverse to the stripe domain pattern (i. e. along the yellow solid arrow in Figs. 3(a-b)). In a large fraction of the sample volume, we observe a typical stripe domain pattern configuration: alternating up/down magnetic domains bounded by Bloch walls with a closure domain structure [36]. This shows up in the transverse cross section of Fig. 3(c) as a circulation of the magnetic moment across the thickness in a series of vortices with opposite circulation sense. However, the characteristic stripe configuration does not extend over the whole sample volume. Near the top, we observe a boundary in the magnetic moment vector map where $m_z$ locally reverses its sign. At the topmost part of the multilayer, $m_z$ contrast becomes much weaker indicating that the magnetic moment is mostly oriented parallel to the sample plane and the stripe pattern is almost suppressed.

The evolution of the amplitude of $m_z$ oscillation across the thickness can be quantified by the average angle of oscillation of the stripe pattern at MD1. As sketched in the inset of Fig. 3(d), $\gamma = \mathrm{asin}(m_z)$, so that the angular oscillation of the stripe pattern can be determined by the average $\langle|\gamma|\rangle$ within each $(x, y, z_i)$ plane, where $z_i$ indicates the plane position through the sample thickness. Figure 3(d) shows a profile of the angle $\langle|\gamma|\rangle$ vs. $z_i$ at MD1. Near the top of the sample, the amplitude of oscillation is about 20°, it decreases to a minimum of 15° at the $m_z$ reversal boundary and reaches a maximum of 55° at a vertical position 120 nm below this boundary. That is, the minimum $\langle|\gamma|\rangle$ occurs near the $GdCo_{TOP}$/NdCo interface whereas its maximum is located near the $GdCo_{BOTTOM}$/NdCo interface, suggesting that the opposite signs of $\varepsilon^{TOP}$ and $\varepsilon^{BOT}$ are responsible for the changes in magnetic moment configuration across the multilayer thickness.

#### 3.1.2 Magnetostatic vs. exchange interactions

We will now rationalize the previous finding on the oscillation amplitudes. First let us consider the central NdCo layer which generates the stripes due to a competition between out-of-plane anisotropy $K_N$ and magnetostatics. Then, there are two energy terms that contribute to imprint the stripe domains into the neighbouring GdCo layers: exchange interaction and magnetostatic coupling (see sketch in Fig. 3(e)). The exchange $E_{ex}$ between Co moments in each layer ($\mathbf{m}(Co)_{GdCo}$ and $\mathbf{m}(Co)_{NdCo}$) can be written as

$$E_{ex} = -J(Co-Co)\mathbf{m}(Co)_{GdCo} \cdot \mathbf{m}(Co)_{NdCo} \qquad (2)$$

and will favour the parallel alignment of $\mathbf{m}(Co)$ at both sides of the interfaces since $J(Co-Co) > 0$.

Next, we consider the magnetostatic energy of the exchange coupled multilayer [36] that, in general, is the sum of two contributions: intralayer energy due to the self-interaction of each magnetic layer and interlayer energy due to the interactions between the layers [37]. In particular, magnetostatic coupling between GdCo and NdCo layers is given by $E_m = -\frac{1}{2}\mu_0 \mathbf{M}(GdCo) \mathbf{H}_d(NdCo)$ [38], i. e. it depends on the interaction between $\mathbf{M}(GdCo)$, the net magnetization at the GdCo layer, and $\mathbf{H}_d(NdCo)$, the field created by magnetic charges at the NdCo layer. Here, it can be written as

$$E_m = -\tfrac{1}{2}\mu_0 \varepsilon \mathbf{m}(Co)_{GdCo} \mathbf{H}_d(NdCo) \qquad (3)$$

where we have explicitly introduced the proportionality factor between $\mathbf{M}(GdCo)$ and $\mathbf{m}(Co)_{GdCo}$. In this way, magnetostatic coupling can be described as an effective field $\varepsilon \mathbf{H}_d(NdCo)$ acting on $\mathbf{m}(Co)_{GdCo}$.

At the GdCo$_{BOTTOM}$/NdCo interface, $\varepsilon^{BOT} \mathbf{H}_d(NdCo)$ and $\mathbf{m}(Co)_{NdCo}$ are parallel (since $\varepsilon^{BOT} > 0$) and with strong out-of-plane components. Thus, they cooperate to imprint the out-of-plane stripe domain oscillation into the bottom GdCo layer as seen in Fig. 3(d). Below the GdCo$_{BOTTOM}$/NdCo interface, the circulating $\varepsilon^{BOT} \mathbf{H}_d(NdCo)$ will create the characteristic closure domain pattern observed in Fig. 3(c) near the bottom of the multilayer to reduce magnetostatic energy of the sample.

The role of this effective field $\varepsilon \mathbf{H}_d(NdCo)$ is qualitatively different at the top GdCo layer. At the GdCo$_{TOP}$/NdCo interface, $\varepsilon^{TOP} \mathbf{H}_d(NdCo)$ is antiparallel to $\mathbf{m}(Co)_{NdCo}$ (since $\varepsilon^{TOP} < 0$) so that magnetostatic and exchange couplings compete with each other, as is typical in ferrimagnetic exchange spring walls [11]. The exchange coupling should decay faster than the magnetostatic one above the interface so that, at some point, $m_z$ will change sign under the effect of $\varepsilon^{TOP} \mathbf{H}_d(NdCo)$ as observed in Figs. 3(a) and 3(b). The vertical location of this $+m_z/-m_z$ boundary is given by the minimum in $\langle \gamma \rangle$ observed in Fig. 3(d).

Thus, the magnetic $+m_z/-m_z$ boundary observed in the magnetic vectorial tomograms in Figure 3 is the signature of a locally modulated exchange spring wall across the multilayer thickness driven by the magnetostatic coupling with the stripe domain pattern of the NdCo layer. The qualitative differences between the magnetic moment configuration near the top and bottom of the sample give further confirmation of the opposite signs of effective interlayer coupling at both GdCo$_{TOP}$/NdCo and GdCo$_{BOTTOM}$/NdCo interfaces in agreement with the multilayer design.

### 3.2 In plane magnetic moments: uniaxial anisotropy vs. interfacial exchange

#### 3.2.1 Analysis of experimental vector tomograms

Figure 4 shows similar views of the vector tomogram at MD1 as discussed above but now contrast is given by the in-plane component of the magnetic moment $m_y$ in order to analyze the in-plane magnetic configuration. Near the top of the multilayer (Fig. 4(a)) we observe an almost uniform $m_y$ contrast, with a smooth oscillation of the magnetic moment around an average in-plane orientation. At the bottom of the multilayer (Fig. 4(b)), the in-plane magnetic moment configuration is less homogeneous and a periodic pattern of closure domains is observed. A cross section of the magnetic moment vector map taken along the solid line in Fig. 4(a)) reveals again qualitative differences in the Gd magnetic moment configuration across the thickness (see Fig. 4(c)). Above the exchange spring boundary, there is an in-plane magnetic domain with almost uniform $m_y$ contrast. Below this boundary, we find the closure domain pattern made of triangular domains with opposite senses of the in-plane magnetic moment transverse to the stripe pattern (see sketch in Fig. 4(e)). However, there are subtle deviations from the standard symmetric closure domains configuration that usually surrounds the Bloch domain walls between stripe domains [32, 36]. Here, cores of closure domain vortices are placed in a zig-zag pattern across the thickness and $-m_y$ closure domains are larger than $+m_y$ domains (see vorticity map in Fig. 4(d), where vorticity is calculated as $|\nabla \times \mathbf{m}|$). This closure domain asymmetry is enhanced near the exchange spring boundary (where the asymmetry in size between $-m_y/+m_y$ domains reaches $\sim 0.2\Lambda$) and implies a non-zero in-plane magnetic moment transverse to the stripe domain pattern [22].

Asymmetries of the in-plane configuration can be quantified in more detail by histograms of the in-plane moment orientations for different reconstructed planes, as shown in the insets of Figs. 4(a-b). Near the top of the multilayer, we find a relatively narrow distribution of in-plane angles centered at the average angle $\varphi_m^{MD1} = -75°$ (calculated from the average of the histogram) within an interval $\Delta\varphi = 60°$ (estimated from the FWHM of the histogram). Two maxima are observed in the angular histogram indicating a smooth ripple oscillation of the in-plane magnetic moment around $\varphi_m^{MD1}$. The aperture $\beta$ of the ripple can be estimated from the distance between these two maxima in the angular distribution as $\beta \approx 35°$. Note that the average magnetic moment orientation in MD1 is not aligned with the stripe pattern orientation (marked by a vertical line at $\varphi_{stripe}^{MD1} = -50°$ at the inset of Fig. 4(a)). Near the bottom of the multilayer, the angular histogram shows two well-spaced maxima ($\beta = 120°$ and $\Delta\varphi = 150°$) corresponding to opposite in-plane closure domains located at symmetric positions relative to the stripe pattern orientation $\varphi_{stripe}^{MD1}$. A certain asymmetry appears between the relative weights of these two maxima in the angular histogram that favours the closure domains nearer to the top layer orientation.

### 3.2.2 Model of competing interfacial exchange and in-plane anisotropy

Several relevant characteristics of the magnetic configuration of the multilayer can be extracted from the analysis of the vector tomogram: a) closure domains are larger for parallel alignment with the magnetic moment at the top MD1; b) the angular distribution of in-plane magnetic moments at the top MD1 is relatively narrow and not far from the stripe pattern orientation; c) the average in-plane magnetic moment transverse to the stripe pattern is non-zero.

These facts suggest a positive effective interlayer coupling between in-plane moments at the GdCo$_{TOP}$/NdCo interface dominated by $J(Co - Co)$ acting on $\mathbf{m}(Co)_{NdCo}$ and $\mathbf{m}(Co)_{GdCo}$ with a negligible contribution from magnetostatic coupling. In-plane demagnetization fields are maximum in the plane transverse to the stripe pattern (and are almost negligible in the direction longitudinal to the stripes), therefore their effect would be very small at the observed $\varphi_m^{MD}$ orientation. Also, the average in-plane moment transverse to the stripes points to a relevant role of in-plane anisotropy not aligned with the stripe domain pattern. In this framework, we propose a simple analytical model to describe the main qualitative features of the multilayer magnetic configuration based in the balance between the interfacial exchange at GdCo$_{TOP}$/NdCo and the in-plane uniaxial anisotropy. The geometry of our model is sketched in Figs. 5(a-b)): at a certain magnetic domain MD, the stripe domain pattern in the NdCo$_5$ layer is oriented at $\varphi_{stripe}^{MD}$, determined by magnetic history, while the in-plane anisotropy axis of the top Gd$_{24}$Co$_{76}$ layer, defined by oblique incidence during sample fabrication, is oriented at $\varphi_K$. The misorientation between the stripe domains and the easy axis can be defined as $\alpha_K = |\varphi_K - \varphi_{stripe}^{MD}|$ taking the shortest angular interval between them.

Now, we assume that this in-plane MD within the top $Gd_{24}Co_{76}$ layer is broken up into a periodic pattern of smaller domains of width 0.5 $\Lambda$, aligned with the underlying closure domain pattern at the top surface of the NdCo layer (see Fig. 5(a)) in order to model the observed ripple oscillation. The orientation of the magnetic moment in these two domains relative to the stripe pattern orientation is given by the in-plane angles $\alpha_1$ and $\alpha_2$. The average orientation of the magnetic moment relative to the stripes direction will be $\alpha_m = \frac{\alpha_1 + \alpha_2}{2}$ and the ripple aperture $\beta = \alpha_1 - \alpha_2$. Then, the anisotropy energy at the top $Gd_{24}Co_{76}$ layer in each stripe period $\Lambda$ can be written as

$$E_K = -\frac{\Lambda}{2} K\big(\cos^2(\alpha_1 - \alpha_K) + \cos^2(\alpha_2 - \alpha_K)\big) \qquad (4)$$

which has the form

$$E_K = -K\Lambda \left(\sin^2\frac{\beta}{2} + \cos\beta \cos^2(\alpha_m - \alpha_K)\right) \qquad (5)$$

in terms of $\beta$ and $\alpha_m$. Thus, the effective anisotropy energy acting on the average magnetic moment of MD oriented at $\alpha_m$ decreases as a function of ripple aperture $\beta$ by a factor $\cos\beta$.

Now, considering only in-plane components of the magnetic moment, we may write the exchange coupling $E_{ex}^{ip}$ at $GdCo_{TOP}$/NdCo interface as

$$E_{ex}^{ip} = -J_{int} \left(d_1 \cos\left(\frac{\pi}{2} - \alpha_1\right) + d_2 \cos\left(-\frac{\pi}{2} - \alpha_2\right)\right) \qquad (6)$$

The first term in eq. (6) corresponds to the exchange between the closure domain of width $d_1$ at the top surface of the NdCo layer with the ripple domain oriented at $\alpha_1$ within the top GdCo layer, whereas the second term describes the exchange interaction between the opposite closure domain of width $d_2$ and the ripple domain oriented at $\alpha_2$. If $\alpha_1$ and $\alpha_2$ are written in terms of the average MD parameters $\alpha_m$ and $\beta$, equation (6) can be expressed as

$$E_{ex}^{ip} = -J_{eff} \Lambda \cos(\alpha_m - \alpha_{ex}) \qquad (7)$$

with $\tan\alpha_{ex} = \frac{d_1 - d_2}{\Lambda} \cotan\frac{\beta}{2}$ and $J_{eff} = J_{int}\sqrt{\frac{(d_1-d_2)^2}{\Lambda^2}\cos^2\frac{\beta}{2} + \sin^2\frac{\beta}{2}}$. Thus, exchange interactions at the $GdCo_{TOP}$/NdCo interface take the form of an effective exchange energy that depends on closure domain asymmetry and on the ripple domain aperture. The sign of the effective exchange angle $\alpha_{ex}$ depends on the closure domain asymmetry $\frac{d_1 - d_2}{\Lambda}$, indicating that $E_{ex}^{ip}$ drives the average in-plane $\mathbf{m}(Co)_{GdCo}$ towards the net in-plane $\mathbf{m}(Co)_{NdCo}$ of the closure domain structure.

The total energy of the average magnetic moment of the MD is a combination of uniaxial anisotropy and effective exchange

$$E_T = -K\Lambda\sin^2\frac{\beta}{2} - K\Lambda \cos\beta \cos^2(\alpha_m - \alpha_K) - J_{eff}\cos(\alpha_m - \alpha_{ex}) \qquad (8)$$

that must be minimized in terms of $\alpha_m$ and $\beta$ for each stripe pattern configuration (given by the closure domain asymmetry $\frac{d_1 - d_2}{\Lambda}$ and the misalignment between the easy anisotropy axis and the stripe pattern orientation $\alpha_K$). In general, uniaxial anisotropy will try to minimize $\beta$ and make $\alpha_m \approx \alpha_K$, whereas exchange interactions will favour large $\beta$ values and $\alpha_m \approx 0$. Figures 5(c-d) show the results of $E_T$ minimization for $J_{int}/K = 1$ and $d_1 - d_2 = 0.2\Lambda$, given by the experimental configuration in Fig.4, as a function of $\alpha_K$.

The evolution of $J_{eff}$ vs. $\alpha_K$ is a consequence of the lateral averaging of exchange interactions, which fluctuate strongly within each period of the stripe pattern, on the larger length scale of a MD modulated by the ripple oscillation around the easy axis. At small $\alpha_K$, i. e. for stripe domains almost aligned with easy anisotropy axis, minimum energy corresponds to $\alpha_m \approx \alpha_K$ and large $\beta \approx 60°$. This results in relatively large $\frac{J_{eff}}{K}$ and small $\alpha_{ex}$, so that the ripple oscillation is symmetric around the stripe pattern and with a large angular aperture. As $\alpha_K$ increases, we observe slight deviations of $\alpha_m$ from the easy axis orientation and a strong reduction of the ripple aperture $\beta$. Effective exchange $J_{eff}$ decreases and the effective exchange angle $\alpha_{ex}$ approaches the orientation of the easy anisotropy axis. Thus, for large $\alpha_K$, the magnetic moment at the top GdCo layer would perform a low amplitude ripple oscillation around the easy anisotropy axis far away from the stripe domain orientation.

It is interesting to mention that a key feature of this model is that $\alpha_K$, the misorientation between the in-plane easy axis and the stripe domains, is the relevant parameter that determines the effective coupling between layers and the global configuration of GdCo layer. $\alpha_K$ can be easily varied with the direction of the last saturating field, providing an external knob to adjust the multilayer configuration with the remanent stripe domain orientation.

### 3.2.3 Global view of in-plane magnetic domains at top GdCo layer

Figure 6 shows a large view of the magnetic tomogram at the top GdCo layer that includes several in-plane MDs with different signs of average $m_y$ contrast, different stripe orientation and different aperture of the ripple oscillations. We will use our analytical model to provide a unified picture of these different MDs and approximately determine the orientation of the in-plane easy axis of the multilayer.

MD1 and MD2 are two domains with similar stripe orientation but opposite signs of average $m_y$ contrast, i. e. opposite signs of their average in-plane $\mathbf{m}(Gd)$ orientation. Angular histograms at MD1 and MD2 (see Fig. 6(b)) show the same qualitative behavior as in Fig. 4: two clear maxima in the angular distribution with an aperture $\beta \approx 35° - 45°$ and average magnetic moment orientation not too far from the stripe pattern orientation, indicated as vertical continuous lines in Fig. 6(b). Note that for each domain there are two possible values of $\varphi_{stripe}^{MD}$, separated 180°, and we have selected the closest one to the measured average orientation for each domain, i.e. $\varphi_{stripe}^{MD1} = -50°$ and $\varphi_{stripe}^{MD2} = 135°$ (instead of the equivalent $\varphi_{stripe}^{MD2} = -45° = 135° - 180°$). With this criterium, the misorientation between the average in-plane magnetization and the stripe pattern is $\alpha_m = |\varphi_m^{MD} - \varphi_{stripe}^{MD}| \approx 25°$, relatively small in both cases, suggesting that $\alpha_K$ is not very large for MD1 and MD2.

Magnetic domains MD3 and MD4 correspond to the other branch of the zig-zag stripe patterns observed in Fig. 2 with stripe orientation $\varphi_{stripe}^{MD4} = +26°$ (and the similar $\varphi_{stripe}^{MD3} = 40° - 180° = -140°$). In this case, the ripple aperture is smaller and, actually, in MD4 a single maximum is observed in the angular distribution centred at $\varphi_m^{MD4} = 91°$. This reduction in ripple aperture and the large angular difference between $\varphi_m^{MD4}$ and stripe pattern orientation, suggest that $\alpha_K$ should be the largest at MD4 and that its average in-plane moment should be oriented along the easy anisotropy axis, i.e. $\varphi_m^{MD4} \approx \varphi_K$ (and, consequently, $\alpha_m \approx \alpha_K$).

We have tested this hypothesis with a comparison between the experimental ripple oscillations observed in the tomogram and $\alpha_1$ and $\alpha_2$ values calculated from the analytical model (with the same parameters used in Fig. 5). The experimental $\alpha_1$ and $\alpha_2$ of the ripple domains have been defined by the positions of the maxima in Figs. 6(b-c) measured as the shortest angular interval from $\varphi_{stripe}^{MD}$ at each MD. $\alpha_K$ is estimated from the shortest $|\varphi_K - \varphi_{stripe}^{MD}|$ interval as indicated in Fig. 6(c) for each of the four MDs, so that we can make a plot of the experimental $\alpha_{1,2}$ $vs.$ $\alpha_K$ values for each particular $\varphi_K$ as shown in Fig. 6(d). The best qualitative agreement is found for $\varphi_K = 95°$, with most experimental data points lying within the angular interval defined by the theoretical $\alpha_1(\alpha_K)$ and $\alpha_2(\alpha_K)$ trends. It corresponds to the easy axis located at the large double arrow in Fig. 6(a), and at the dashed vertical lines at the histograms in Figs. 6(b-c). In all cases, the average magnetization orientation lies close to the easy axis with a slight displacement towards the stripe domain orientation, in line with the predictions from the analytical model based on the lateral averaging of exchange interactions within the ripple pattern. That is, our multilayer is an experimental realization of an exchange spring system in which effective coupling is not simply determined by material parameters fixed during sample fabrication but, rather, it can be externally adjusted with magnetic history (i. e. with the orientation of the stripe pattern at remanence).

### *3.3 Domain wall between chevron magnetic domains*

Finally, let us discuss the magnetic configuration at the corners of the stripe domain chevron pattern (e.g. at the boundary between MD1 and MD4 as sketched in Fig. 7(a)). Here, the stripe pattern rotates by $\Delta\varphi_{stripe}^{MD1-MD4} = 76°$ whereas the average magnetic moment orientation changes by $\Delta\varphi_m^{MD1-MD4} = 166°$ creating a high angle domain wall between MD1 and MD4. This wall is oriented at $\varphi_{DW} \approx 82°$, i.e. it is almost parallel to the average in-plane magnetic moment at the top GdCo layer of MD1 and MD4 and makes similar angles with the stripe patterns in both chevron MDs so that it is a low energy wall from the magnetostatic point of view.

The domain wall has a complex 3D configuration with a peculiar undulating pattern of period $\Lambda$ of interleaving $\pm m_y$ regions. Also, the vertical cross section along the domain wall core (Figs. 7(b-c)) presents clear differences from the configuration within each MD (shown in Fig. 4).

First, note that the size of $+m_y/-m_y$ closure domains is similar and closure vortex cores are aligned at a constant $z_i$ position within the multilayer. This can be attributed to the continuous transition between MD1 and MD4. Within each MD, the sign of the average in plane magnetic component $\langle m_y \rangle = \sin(\varphi_m^{MD})$ is directly related with the sign of $\delta d = d_1 - d_2$, the closure domain asymmetry. Since $\langle m_y \rangle$ and, consequently, $\delta d$ have opposite signs at MD1 and MD4, $\delta d$ must become zero at the domain wall, resulting in the same size $+m_y/-m_y$ closure patterns observed in Fig. 7(b).

Second, at the exchange spring wall near the top of the multilayer, there is a sign reversal in both $m_y$ and $m_z$ components (within each MD the reversal takes place only in $m_z$). This indicates that the dominant coupling term at the GdCo$_{TOP}$/NdCo interface favours an antiparallel alignment both for in-plane and out-of-plane magnetic moments (within each MD antiparallel coupling dominates only in $m_z$).

These observations can be attributed to a subtle change in the competition between the relevant energy terms at the domain wall driven by the rotation of the stripe pattern at the corner of the zig-zag: as the stripe pattern turns from $\varphi_{stripe}^{MD1} = -50°$ to $\varphi_{stripe}^{MD4} = +26°$ it becomes perpendicular to the in plane anisotropy easy axis. This fact has two consequences: 1) in-plane anisotropy axis becomes aligned with the closure domain pattern, 2) in-plane components of $H_d$ are strongest at the orientation favoured by in-plane anisotropy. Thus, in-plane anisotropy enhances the in-plane antiparallel coupling term proportional to the effective field $\varepsilon^{TOP}H_d$ resulting in the sign reversal of $m_y$ at the exchange spring wall. The enhancement of in-plane antiparallel coupling allows the closure domain pattern to emerge at the top GdCo layer and creates the undulating domain wall profile seen in Fig. 7(a).

At the intersection between the exchange spring wall and the closure domain pattern, the $\mathbf{m}(Gd)$ configuration is highly disordered and a high density of vortices and antivortices appears to accommodate these different competing interactions. For example, several antivortices across the thickness are observed at the tip of out-of-plane domains whereas at other locations the system has nucleated closely spaced vortex/antivortex pairs (see Figs. 7(d-e)).

## 4. Conclusions

A $Gd_{12}Co_{88}/Nd_{17}Co_{83}/Gd_{24}Co_{76}$ multilayer has been prepared with competing magnetostatic, exchange and anisotropy interactions and its 3D magnetic configuration has been determined from experimental data by X-ray magnetic vector tomography at the Gd M absorption edges.

The general view of the magnetization at MTXM images shows stripe domains with alternating up and down magnetization arranged in a chevron type of pattern with alternate positive and negative in-plane magnetizations.

The vector tomogram reveals a strong out-of-plane magnetization oscillation at the bottom GdCo layer with a closure domain structure across the thickness. It is caused by the cooperation between ferromagnetic exchange and magnetostatic coupling at the GdCo$_{BOTTOM}$/NdCo interface that imprints stripe domains into the GdCo layer and, also, by the need to reduce magnetostatic energy at the multilayer/vacuum interface.

The magnetization of the top GdCo layer is mostly in-plane, with a weak out-of-plane oscillation, and medium sized in-plane MDs that extend over the different sections of the chevron zig-zags. Local exchange springs, characterized by the sign reversal of $m_z$ across the thickness, are observed near the GdCo$_{TOP}$/NdCo interface as a result of the effective antiparallel magnetostatic coupling between NdCo and top GdCo.

Ripple patterns within chevron-like domains are observed at the top GdCo layer, separated by undulating domain walls decorated by magnetic vortices and antivortices across the thickness. They are the result of the subtle balance between effective antiparallel magnetostatic coupling (which is maximum perpendicular to the stripe domains and minimum along them), uniaxial anisotropy defined during the sample fabrication process and interfacial ferromagnetic exchange (isotropic).

In summary, the detailed description of the trilayer magnetization provided by vector magnetic tomography has allowed us to unravel experimentally the complex behavior of the system without a priori assumptions. It is found that the relevant parameters to obtain an adjustable exchange spring are the fine tuning of interlayer coupling by GdCo stoichiometry, defined during sample fabrication, and the orientation of the magnetic stripes relative to in-plane uniaxial anisotropy, adjustable by magnetic history.

## Acknowledgements

Alba light source is funded by the Ministry of Research and Innovation of Spain, by the Generalitat de Catalunya and by European FEDER funds. This project has been supported by Spanish MICINN under grant PID2019-104604RB/AEI/10.13039/501100011033.

## Methods

*Sample Fabrication and Magnetic characterization.* $Gd_{12}Co_{88}/Nd_{17}Co_{83}/Gd_{24}Co_{76}$ multilayers to create exchange springs at the top $Gd_{24}Co_{76}$ layer were grown by sputtering on 50 nm thick $Si_3N_4$ membranes as reported before [23, 39]. In-plane uniaxial anisotropy of top/bottom GdCo layers was defined during sample deposition by the oblique projection of the direction of the atomic beams relative to the sample surface [40]. The easy axis orientations are parallel at top/bottom layers. Saturation magnetization and uniaxial anisotropy at room temperature in each layer have been obtained from Vibrating Sample Magnetometry (VSM) and MOTKE hysteresis loops [11]: $M_S(GdCo)^{BOT} = 5.1 \times 10^5$ A/m, $K(GdCo)^{BOT} = 5.1 \times 10^3$ J/m³ and $M_S(GdCo)^{TOP} = 8.94 \times 10^4$ A/m, $K(GdCo)^{TOP} = 1.34 \times 10^4$ J/m³. At the central NdCo₅ layer, typical parameters are $M_S(NdCo) = 7 \times 10^5$ A/m and out of plane anisotropy $K_N(NdCo) \approx 10^5$ J/m³ [25], which is an order of magnitude larger than in GdCo layers. Thus, the NdCo₅ layer with its stripe domain pattern will provide the pinned magnetic configuration within the multilayer.

The 80 nm $Gd_{12}Co_{88}$/80 nm $Nd_{17}Co_{83}$/80 nm $Gd_{24}Co_{76}$ multilayer was prepared in a multidomain state after out-of-plane demagnetization allowing the system to explore the complex energy landscape created by the competing interactions and observe it in a single experiment.

*MTXM and soft X-ray vector tomography.* The sample was mounted in the high precision rotary stage of the full field X-ray transmission microscope at the MISTRAL beamline of ALBA Synchrotron [32]. 100 nm gold nanoparticles dispersed on its surface serve as fiducials accurate projection alignment to a common rotation axis prior to the tomography reconstruction. It was illuminated with circularly polarized X-rays with fixed polarization in order to exploit magnetic contrast from X-ray circular dichroism [41] and it was rotated around an axis parallel to the sample surface and perpendicular to the horizontal X ray beam, in order to acquire a tilt series of images, i.e. a set of closely spaced MTXM images at 2° intervals in the angular range $\vartheta = \pm 26°$ and at 1° intervals in the angular ranges $[-55^o, -26^o]$ and $[26^o, 55^o]$ (see sketch in Fig. 2, $\vartheta = 0°$ when the multilayer normal is parallel to the X ray direction). Tomographic reconstruction of the magnetization vector requires two orthogonal tilt series [29, 30]. Thus, the sample was manually rotated in-plane by an angle $\phi \approx 90°$ and a second tilt series was acquired. Fine alignment of the two tilt series together was also performed as a part of the tomographic reconstruction process with the aid of the gold fiducials [32]. The actual in-plane angle between the two-tilt series was measured as $\phi \approx 95°$. At each $(\vartheta, \phi)$ orientation, 2D transmitted images were sequentially acquired at the Gd $M_4$ (wavelength 1.0198 nm) and Gd $M_5$ (wavelength 1.0456 nm) absorption edges since they present opposite signs of the X-ray magnetic circular dichroic factor. Then, either charge (TXM) or magnetic (MTXM) contrast images were obtained by the addition/subtraction of the logarithm of individual transmittance images at Gd $M_4$ and $M_5$ edges with a proper normalization to minimize magnetic contrast in the charge images [32, 42]. In general, magnetic contrast is given by the projection of the magnetic moment along the X-ray beam direction [41]. Normal incidence images are only sensitive to the out-of-plane component of the magnetic moment $m_z$ whereas tilted incidence images also provide information of the in-plane component of the magnetic moment perpendicular to the rotation axis. Thus, images in Tilt series 1 are sensitive to $m_y$ and $m_z$, whereas images in Tilt series 2 are sensitive to $m_z$ and approximately $m_x$.

We can obtain the 3D magnetic moment configuration in the multilayer from the reconstruction of the MTXM data sets in Tilt series 1 and 2 with a tomography algorithm using the method reported in [29, 32]. The total reconstructed volume is $2700 \times 2700 \times 2025$ nm$^3$ (with a total of $200 \times 200 \times 150$ voxels), which is much larger than the multilayer thickness. The tomography algorithm properly locates, without any *a priori* assumptions, the magnetic signal from the multilayer just below the plane defined by the gold fiducials that were placed on the sample surface. The result is a 3D vector map of the magnetic moments of Gd, $\mathbf{m}(Gd)$, convoluted with the lateral resolution function of the microscope (~30 nm) and the axial resolution of the measurement. The latter has been estimated to be of ~60 nm which is comparable with individual layer thickness [32]. An important methodological issue that has to be clear is the following: as we do not impose any *a priori* condition (i.e., we do not set $m = 0$ at the NdCo central layer), our tomographic reconstruction generates a continuous field of magnetic moments that, due to the finite axial resolution is non zero in the central layer in spite of not having Gd atoms. This continuity mimics the effect of exchange coupling between layers [30]. Here, we have focused on the differences in configuration between the upper and lower regions of the multilayer (i.e., GdCo layers), with a more qualitative discussion of the evolution across the thickness.

**Data availability**
The datasets generated during the current study are available from the corresponding author on reasonable request

**Competing interests**

The authors declare that they have no competing financial interests.

**Author contributions**
C.Q., J. I. M., M.V. and S.F. designed and planned the experiments, J. H., C.Q. and J.I.M. prepared the samples. J. H., A.H.-R., C.Q., A.S., S.R., E.P. and S.F carried out the synchrotron experiment. J. H. and A.H.-R. performed the tomographic reconstructions. J. H., A.H.-R., L.M.A.P., J.I.M., C.Q., M.V. and S.F. analyzed the data. All the authors participated in the interpretation and discussion of the results. A.H.-R., M.V. and S.F. wrote the article.

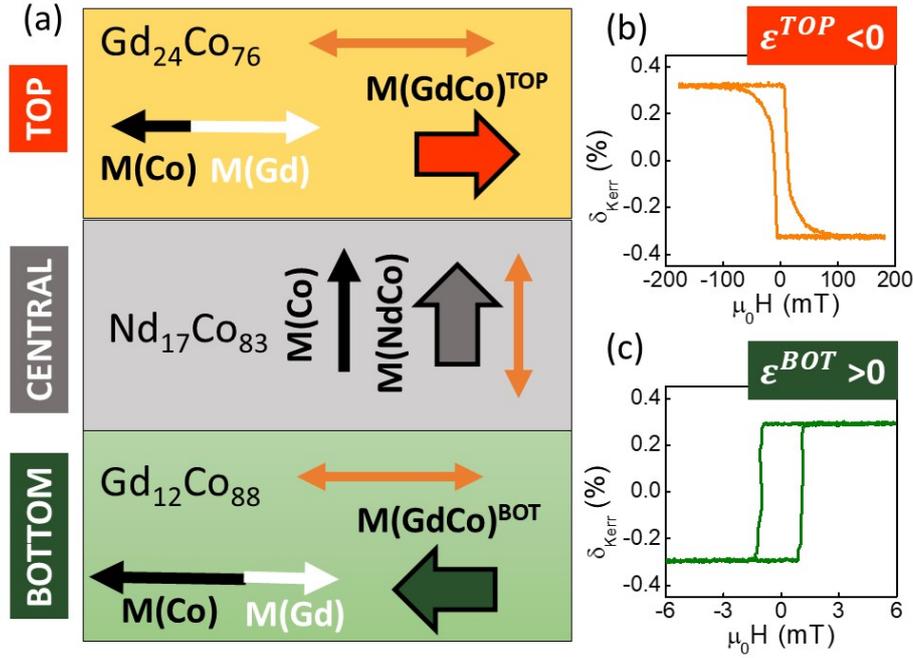

**Figure 1:** *(a) Sketch of multilayer structure. Black/white thin arrows indicate magnetization of Co and Gd atom sublattices, $\mathbf{M}(Co)$ and $\mathbf{M}(Gd)$, respectively. Thick arrows indicate net magnetization $\mathbf{M}(GdCo)$. Double orange arrows indicate anisotropy easy axis. (b) Room temperature MOTKE hysteresis loop of 50 nm $Gd_{24}Co_{76}$ control sample, sensitive only to Co atom sublattice. Note the negative signal at saturation in a positive field, indicating the antiparallel alignment between $\mathbf{M}(GdCo)^{TOP}$ and $\mathbf{m}(Co)^{TOP}$, i.e. $\varepsilon^{TOP} < 0$. (c) Room temperature MOTKE hysteresis loop of 50 nm $Gd_{12}Co_{88}$ control sample with positive signal at positive saturation field, corresponding to parallel $\mathbf{M}(GdCo)^{BOT}$ and $\mathbf{m}(Co)^{BOT}$, i.e. $\varepsilon^{BOT} > 0$.*

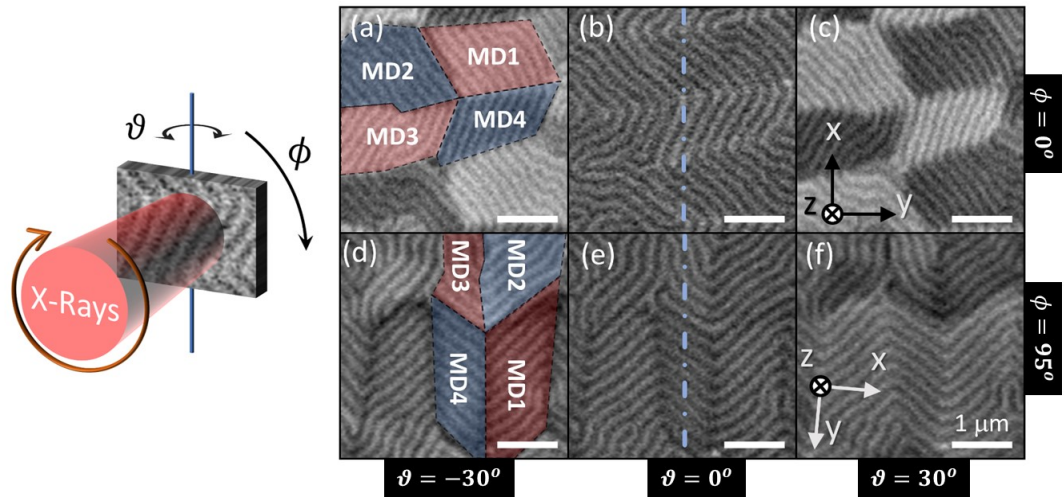

**Figure 2: MTXM images at different angular orientations at Gd M edge.** *Sketch of X-ray beam/sample geometry at the MISTRAL microscope indicating angles $\vartheta$ and $\phi$.* <u>*Tilt series 1 ($\phi = \underline{0°}$):*</u> *a) $\vartheta = -30°$; b) $\vartheta = 0°$; c) $\vartheta = +30°$.* <u>*Tilt series 2 ($\phi = 95°$):*</u> *d) $\vartheta = -30°$; e) $\vartheta = 0°$; f) $\vartheta = +30°$. Dashed line indicates rotation axis for each tilt series. Labels indicate Magnetic Domains (MD) with different average in-plane magnetization.*

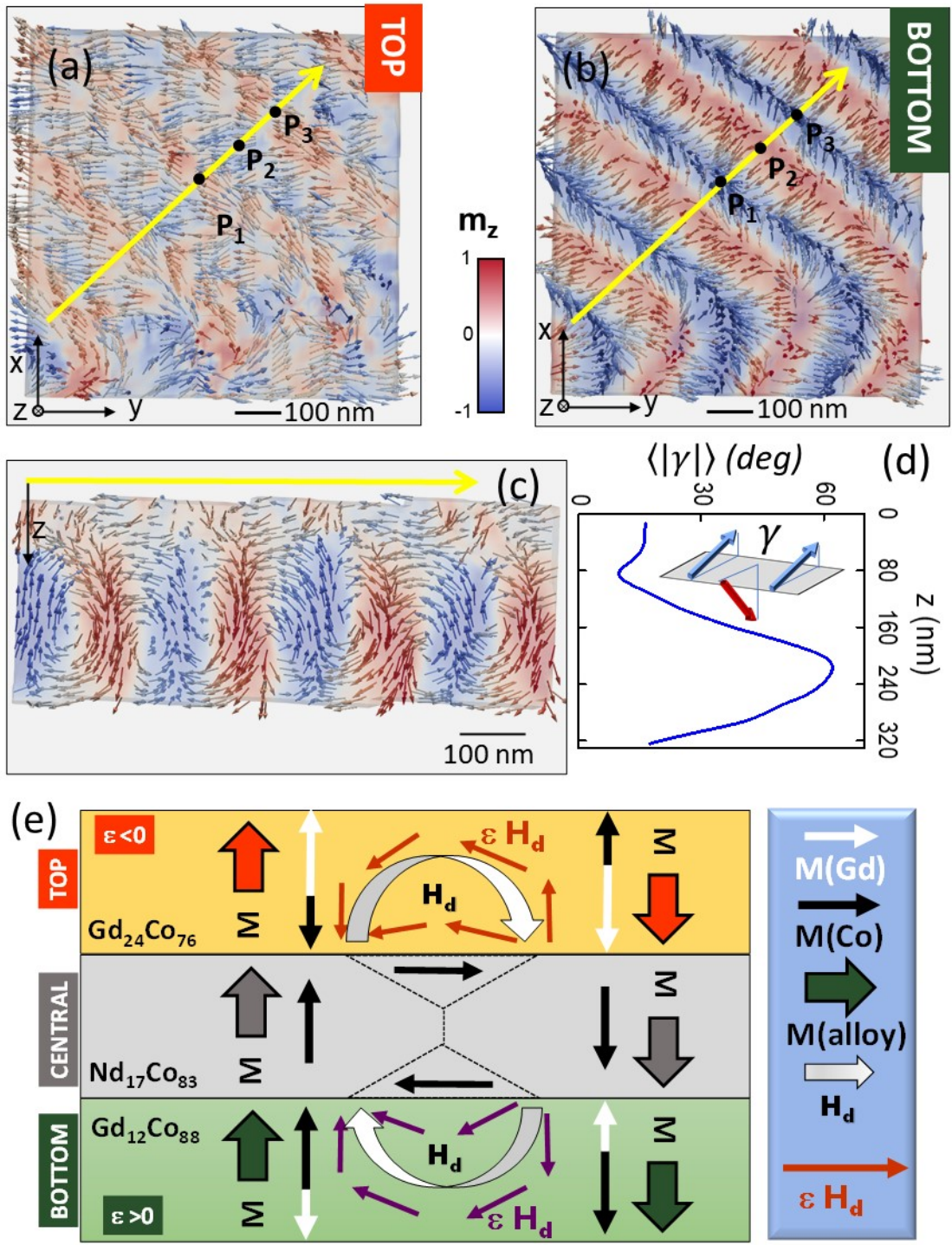

**Figure 3: Vectorial tomographic reconstruction of $\mathbf{m}(Gd)$ at MD1 ($m_z$ contrast).** *(a) View of the $(x, y, 14\ nm)$ plane (Top Gd-Co layer); (b) View of the $(x, y, 255\ nm)$ (bottom Gd-Co layer); (c) Cross-section across yellow arrow in (a). Note the reversed $m_z$ contrast across the sample thickness; (d) Profile of average magnetization oscillation across the thickness (measured as $\langle|\gamma|\rangle$ within each $(x, y, z_i)$ plane). Inset shows a sketch of out-of-plane magnetic moment oscillation within a $(x, y, z_0)$ plane with amplitude $\gamma$; (e) Sketch of multilayer configuration and out-of-plane competing interactions.*

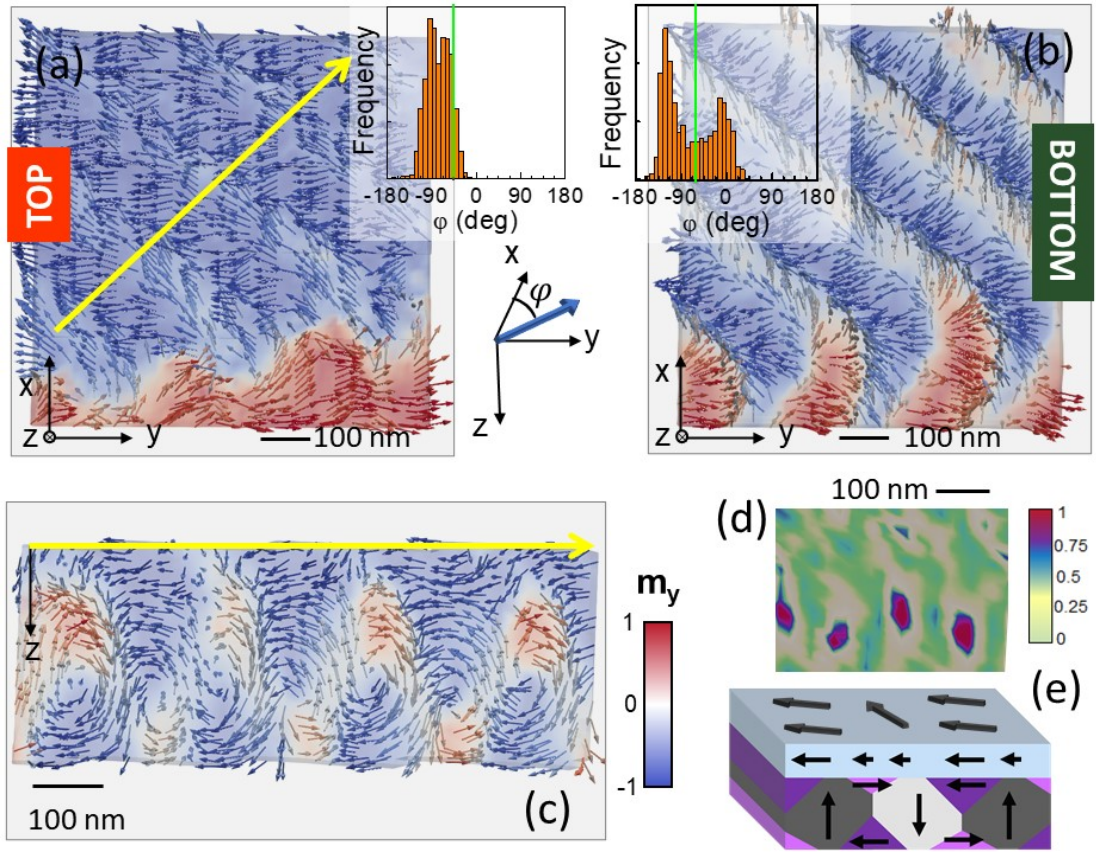

**Figure 4: Vectorial tomographic reconstruction of $\mathbf{m}(Gd)$ at MD1 ($m_y$ contrast).** *(a) View of the $(x, y, 14\ nm)$ plane (Top Gd-Co layer); (b) View of the $(x, y, 255\ nm)$ plane (bottom Gd-Co layer); Insets are histograms of in-plane angular orientations respect to the $x$-axis ($\varphi$) within top/bottom layers respectively. Vertical line marks stripe domain orientation ($\varphi_{stripe}^{MD1} = -50°$). (c) Cross-section across yellow arrow in (a). Note the uniform $m_y$ contrast at top layer and the asymmetric size of opposite closure domains. (d) Detail of the vorticity map at cross section in (c). Note the zig-zag configuration of high vorticity regions corresponding to cores of closure domain vortices. (e) Sketch of domain configuration across the thickness*

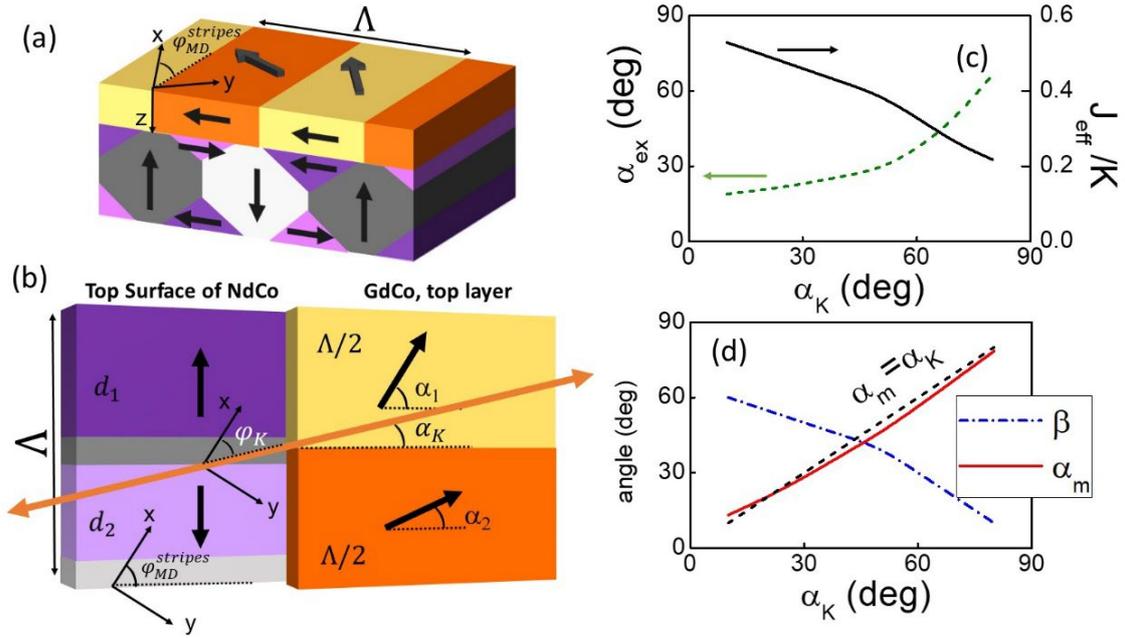

***Figure 5: Model of in-plane interactions**. a) Sketch of cross section of the multilayers with asymmetric closure domains. (b) Sketch of relevant exchange interactions between asymmetric closure domains (width $d_1$ and $d_2$) at the top surface of the NdCo layer and ripple domains at GdCo top layer (oriented at $\alpha_1$ and $\alpha_2$). Double arrow indicates in-plane easy axis at top GdCo layer oriented at an angle $\alpha_K$ relative to stripe domain pattern. (c) Dependence of effective exchange energy $J_{eff}/K$ and exchange angle $\alpha_{ex}$ on $\alpha_K$ for $J_{int}/K=1$ and $d_1 - d_2 = 0.2\ \Lambda$. (d) Dependence of average magnetization orientation $\alpha_m$ and ripple amplitude $\beta$ on $\alpha_K$. Dashed black line corresponds to the condition of symmetric ripple around easy anisotropy axis.*

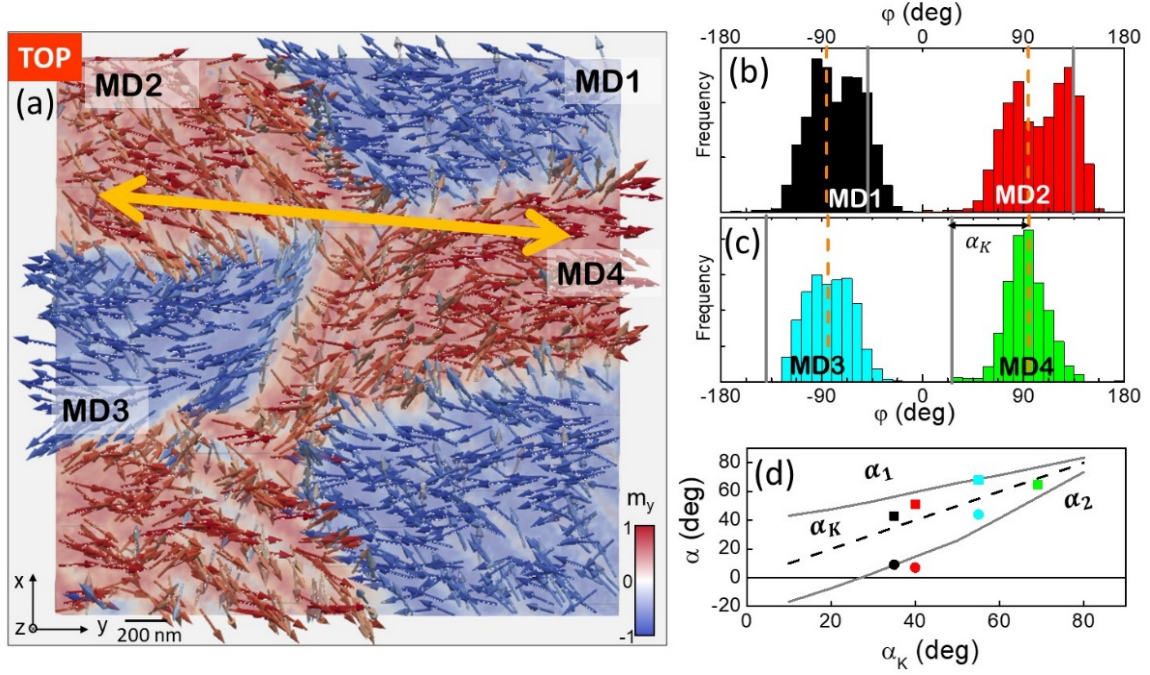

**Figure 6: Statistics of magnetic moment orientation in the ripple states.** *(a) Vectorial tomogram reconstruction of* $\mathbf{m}(Gd)$ *at the* $(x, y, 14\ nm)$ *plane (Top Gd-Co layer) with* $m_y$ *contrast. Double arrow indicates the easy anisotropy axis derived from the analytical model. (b-c) Histograms of in-plane angular orientations* $(\varphi)$ *respect to the x-axis for the different MDs. Vertical solid lines indicate stripe orientation* $\varphi_{stripe}^{MD}$ *in each MD region and vertical dashed lines indicate easy axis orientation* $\varphi_K$. *(d) Ripple domain orientation* $\alpha_1$ *and* $\alpha_2$ *vs.* $\alpha_K$. *Solid lines are calculated from the analytical model with the same parameters as in Fig. 5. Symbols correspond to experimental* $\alpha_1$ *and* $\alpha_2$ *obtained from the shortest angular distance between maxima in the histograms and* $\varphi_{stripe}^{MD}$. *The parameter* $\alpha_K$ *in each MD is obtained from the shortest angular distance between* $\varphi_{stripe}^{MD}$ *and* $\varphi_K$ *as indicated in (c).*

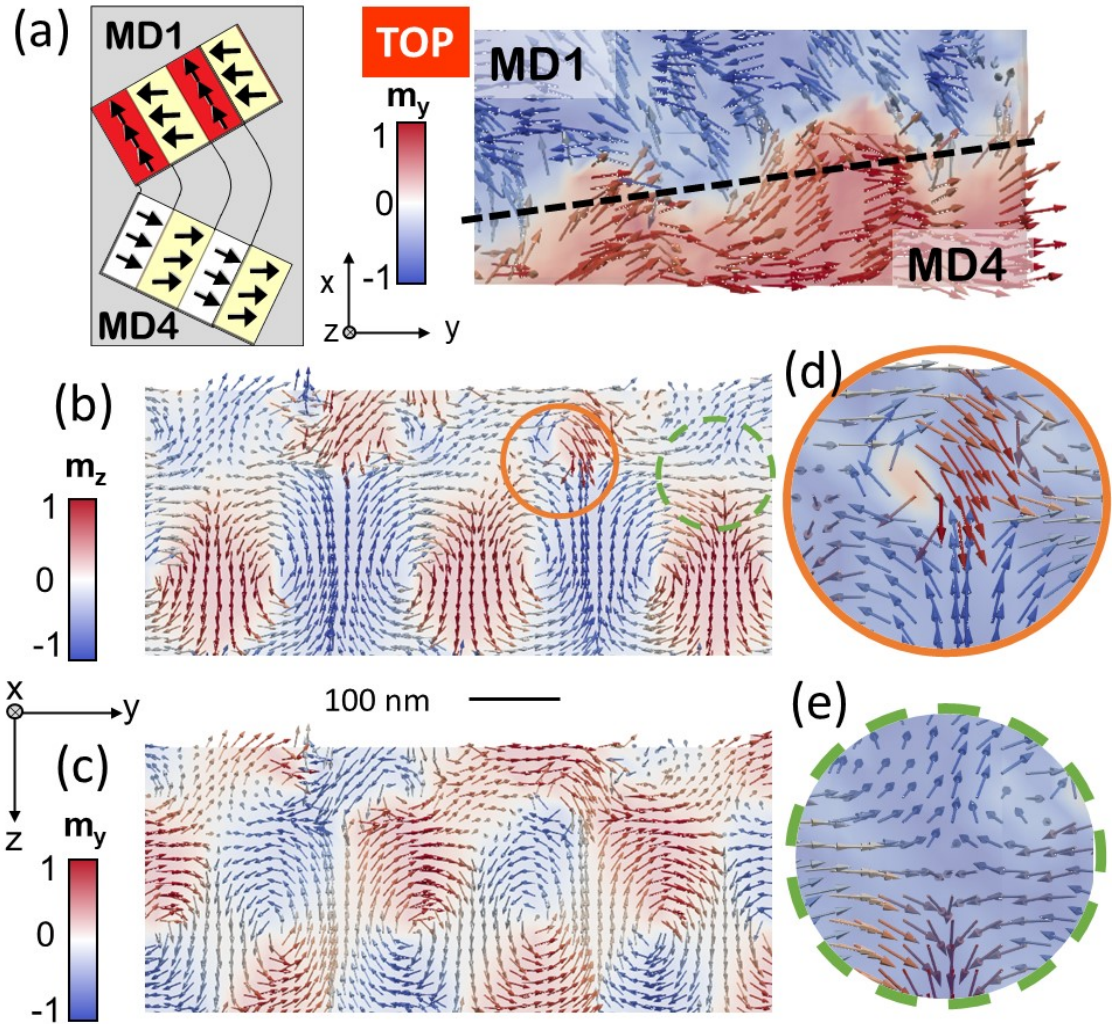

***Figure 7: Tomographic reconstruction of $\mathbf{m}(Gd)$ at the domain wall between MD1 and MD4.***
*a) Sketch of the boundary between MD1 and MD4 at a zig-zag corner of the stripe pattern and zoom view of magnetic tomogram at the $(x, y, 14\,nm)$ plane (Top Gd-Co layer) with $m_y$ contrast. (b-c) Vertical cross-section along domain wall (dashed line in (a)) with $m_z$ and $m_y$ contrast, respectively. Note the antiparallel alignment of the magnetic moments at the exchange spring wall across the thickness both in $m_z$ and $m_y$ components and the symmetric configuration of closure domains. (d-e) Details of localized vortices (orange circle in (b)) and antivortices (green circle in (b)) at the interface.*